\documentclass[english,preprint]{revtex4-1}
\usepackage[T1]{fontenc}
\usepackage[latin9]{inputenc}
\usepackage{xcolor}
\usepackage{pdfcolmk}
\usepackage{bm}
\usepackage{amsbsy}
\usepackage{graphicx}
\usepackage{esint}
\PassOptionsToPackage{normalem}{ulem}
\usepackage{ulem}

\makeatletter

\providecolor{lyxadded}{rgb}{0,0,1}
\providecolor{lyxdeleted}{rgb}{1,0,0}

\DeclareRobustCommand{\lyxsout}[1]{\ifx\\#1\else\sout{#1}\fi}

\makeatother

\usepackage{babel}
\begin{document}
\title{Microturbulence-mediated route for energetic ion transport and Alfvénic
mode intermittency in tokamaks}
\author{N. N. Gorelenkov$^{\dagger}$ and V. N. Duarte$^{\dagger}$}
\affiliation{$^{\dagger}$\emph{Princeton Plasma Physics Laboratory, Princeton
University, Princeton, NJ, 08543, USA}}
\begin{abstract}
We report on a theoretical discovery of new regimes of Alfvén eigenmode
(AE) induced fast ion transport in tokamak plasmas, where microturbulence
plays the role of a mediator of fast ion relaxation. Coulomb collisional
scattering alone leads to small AE amplitudes and does not reproduce
the steady state regimes observed in experiments. We show that in
nonlinear regimes the effective pitch angle scattering due to microturbulence
can lead to steady state AE amplitude evolution. This indicates a
new route for fast ion losses, which is beyond the scenarios described
in ``Energetic ion transport by microturbulence is insignificant
in tokamaks'' {[}D. C. Pace et al., Phys. Plasmas \textbf{20} (2013)
056108{]}. As a result, microturbulence can significantly increase
the amplitude of AEs in predictive simulations of burning plasma experiments
such as ITER.
\end{abstract}
\maketitle
\textcolor{black}{The success of the next generation of fusion devices
relies on their ability to confine fusion alpha particle products
long enough to transfer a substantial fraction of their energies to
the reacting thermal ions. The International Thermonuclear Experimental
Reactor (ITER) is predicted to have tight tolerance for fast ion losses
in order to sustain burning plasmas \citep{FasoliNF07}. ITER is expected
to have a multitude of unstable modes in the toroidal Alfvén eigenmode
(TAE) frequency range \citep{FitzgeraldNF16,SchnellerPPCF16} and,
therefore, can likely lead to more global losses via resonance overlapping.
It is essential, therefore, to develop efficient and robust capabilities
to predict the relaxation of energetic ion component in tokamak experiments.
The effect of the pitch angle scattering due to the microturbulence
on the saturation amplitudes has previously been ignored, which leads
to a significant underestimation of the level of }activity of AEs.

\textcolor{black}{This Letter shows that in fusion plasmas, the scattering
frequency, $\nu_{\chi\chi}$, of} fast ion pitch angle, $\chi=v_{\|}/v$,
due to Coulomb collisions is too small to bring the unstable AEs to
steady state regimes observed in experiments \citep{CollinsPRL16}.
It has been noted earlier that Coulomb scattering is not sufficient
to explain the observed AE amplitudes in TFTR when the ion cyclotron
resonance heating (ICRH) was applied \citep{WongPoP97}. In those
experiments, strong scattering was required and was shown to be the
result of applied ICRH. Subsequent analysis using a cubic amplitude
evolution equation \citep{BerkPRL96} helped to describe the nonlinear
saturation of $n=2$ TAE at $\sim10$ times higher amplitude, as well
as helped to evaluate the growth rate of the unstable mode.

By itself the effect of pitch angle scattering is not new and was
reproduced in many publications, including more recent ones, such
as \citep{SlabyNF18} and \citep{GorelenkovPoP19}. If multiple unstable
AEs are mediated by the microturbulence, the induced scattering can
play a profound role in energetic particle (EP) relaxation. Namely,
the additional scattering frequency can strongly enhance the AE amplitudes
and drive them into steady state regimes that in turn account for
substantial fast ion losses. Ref. \citep{Pace2013} considered the
effects of microturbulence and Alfvén waves on fast ion transport
separately and concluded that the effect of the microturbulence with
or without AEs is weak and does not lead to a significant experimentally
observable EP transport. We show with the example of a single mode
that even though the direct effect of the microturbulence is small
for fast ion transport, together with AEs it offers a new route for
fast ion radial transport and losses by enlarging the resonance extent,
and thereby, by boosting the amplitude of each eigenmode. We show
that two effects are indissociable since the turbulence strength is
key in setting the amplitudes. The amplitudes of AEs, in turn, set
the EP radial transport and losses. This leads to the conclusion that
the additional scattering mechanism is required, in concert with the
microturbulence scattering considered in the development and validation
of a criterion for whether AEs should exhibit a chirping or a quasi-steady
frequency response \citep{DuarteNF17,DuartePoP17}.

The pitch angle scattering self-consistently enters the quasilinear
(QL) methodology via the second order differential scattering operator
acting on the EP distribution function and via the broadening of the
resonance layer \citep{DuartePoP19}, as discussed later. As such,
the additional scattering is crucial for QL model to work properly.
Most of the initial value codes, such as those recently benchmarked
in linear regimes in Ref. \citep{Taimourzadeh2019} as well as recent
ITER projection models \textcolor{black}{\citep{FitzgeraldNF16,SchnellerPPCF16}},
employ a scattering operator ignoring contributions from microturbulence.

\label{Recent-experiments-have}Recent experiments have shown the
resilience of EP profiles to the neutral beam injection (NBI) power
\citep{CollinsPRL16}. Those experiments showed the transport regulated
by stochasticity and therefore the anomalous scattering can be the
key in understanding the dynamics. More than ten Alfvénic modes with
low amplitudes $\delta B_{\theta}/B\sim O\left(10^{-4}-10^{-2}\right)$
were excited in steady state regimes during the experiments, which
produced a noticeable effect on EP confinement setting up ``stiff''
density profiles. We should note that even though we discuss a single
mode saturation behavior our conclusions are important for multiple
instabilities which were shown to be relevant for EP relaxation \citep{BerkNF95}
although more detailed analysis is beyond this paper goals. Correct
diffusion representation due to a single mode in the nonlinear regime
is important even for the case when AE modes and their resonances
overlap.

Here we apply two codes. The first one is the Resonance Broadened
Quasilinear (RBQ) code validated for near-threshold conditions \citep{GorelenkovPoP19}
expected in experiments \citep{CollinsPRL16}. And, the second code
is the kinetic simulation code BOT which solves the relaxation of
a bump-on-tail distribution function in a model formulation \citep{Lilley2010,LilleyBOT14}.

\textbf{\textit{Formulation}}\textbf{. }The RBQ code has been built
with the goal of efficiently computing the relaxation of EP distribution
function in the presence of multiple AEs \citep{GorelenkovPoP19,DuartePhD17}.
It implements the QL equations to compute the EP distribution function
and relaxes it in time along the canonical toroidal momentum $P_{\varphi}$
according to the equation which sums the diffusion operator of all
the modes under consideration: 
\begin{equation}
\frac{\partial f}{\partial t}=\sum_{k,p,m,m'}\frac{\partial}{\partial P_{\varphi}}D_{kp}(P_{\varphi};t)\frac{\partial}{\partial P_{\varphi}}f+\left\langle \frac{1}{r}\frac{\partial}{\partial r}D_{rh}r\frac{\partial(f-f_{0})}{\partial r}\right\rangle +\left\langle \frac{\partial}{\partial\chi}\nu_{\chi\chi}\frac{\partial(f-f_{0})}{\partial\chi}\right\rangle ,\label{eq:RBQDiffOp}
\end{equation}
where the diffusion coefficients due to AEs are expressed as $D_{kp}(P_{\varphi};t)=\pi C_{k}^{2}\left(t\right)\mathcal{E}^{2}\mathcal{R}_{\bm{l}}G_{km'p}^{*}G_{kmp}$,
$\mathcal{R}_{\bm{l}}=\mathcal{R}_{\bm{l}}\left(P_{\varphi}-P_{\varphi r}\right)$
is the resonance window function analytically formulated in Ref. \citep{DuartePoP19},
$G_{kmp}$ are the wave-particle interaction matrix elements of the
$k$-th mode of amplitude $C_{k}$, $m$-th poloidal harmonic, $p$-th
resonant side-band, the resonance center is given by the condition
$P_{\varphi}=P_{\varphi r}$, $\chi=v_{\|}/v$ is the pitch angle,
and $\nu_{\chi\chi}=\nu_{\perp}\left(1-\chi^{2}\right)$ for the case
of Coulomb collisions with $\nu_{\perp}$ being the $90^{o}$ scattering
frequency \citep{GoldstonJCP81}, $D_{rh}$ is the radial diffusion
coefficient of energetic or hot ions due to the microturbulence \citep{LangFu2011}
and $\left\langle ...\right\rangle $ denotes orbit averaging. The
second order derivative terms on the RHS of Eq.(\ref{eq:RBQDiffOp})
are the terms which need to be kept near the resonances since they
are responsible for the variation of the distribution function $f$
in their vicinity where $f$ deviates the most from the initial equilibrium
distribution function, $f_{0}$. The time scale of resonant particle
dynamics near the resonance is very short, on the order of $0.1-0.5~msec$,
which is much shorter than the injection time or slowing down scale,
which is $20-100msec$. Such time scale separation is sufficient to
describe the problems prescribing the EP flux intermittency.

\label{Ckeqs}Equation (\ref{eq:RBQDiffOp}) is supplemented by the
equation for AE amplitudes $dC_{k}^{2}\left(t\right)/dt=2\left(\gamma_{L,k}+\gamma_{d,k}\right)C_{k}^{2}\left(t\right)$,
where local growth rates, $\gamma_{L,k}=\gamma_{L,k}\left(t\right)$,
are computed at each time $t$ using the distribution function $f$
although the damping rate is fixed in time.

In tokamaks, the effect of microturbulence on the EP relaxation can
be projected onto one direction with good accuracy when $n\gg1$ and
$\omega\ll\omega_{c}$, where $\omega_{c}$ is the cyclotron frequency.
This is often the case in experiments. Since the canonical momentum
is linearly proportional to $\chi$ and the poloidal magnetic flux
function, $\psi$, $P_{\varphi}=e\psi/2\pi mc-\chi vRB_{\varphi}/B,$
one finds that $dP_{\varphi}\bigr|_{\psi,\mathcal{E}}=-\left(vRB_{\varphi}/B\right)d\chi$
and $dP_{\varphi}\bigr|_{\mathcal{E},\chi}=\frac{e}{2\pi mc}d\psi$.
From Eq.(\ref{eq:RBQDiffOp}) it follows that the last two terms on
its RHS contribute to the diffusion in $P_{\varphi}$ direction. Then
combining them reduces Eq.(\ref{eq:RBQDiffOp}) to

\begin{equation}
\frac{\partial f}{\partial t}\simeq\sum_{k,p,m,m'}\frac{\partial}{\partial P_{\varphi}}D_{kp}(P_{\varphi};t)\frac{\partial}{\partial P_{\varphi}}f+\left\langle \left[R_{Dh}+1\right]\frac{\partial P_{\varphi}}{\partial\chi}\frac{\partial}{\partial P_{\varphi}}\nu_{\chi\chi}\frac{\partial P_{\varphi}}{\partial\chi}\right\rangle \frac{\partial(f-f_{0})}{\partial P_{\varphi}},\label{eq:RBQDiff1D}
\end{equation}
where the ratio of hot particle diffusion rates in $P_{\varphi}$
direction due to the turbulence and collisions can be written for
the case of weak radial dependence of $\nu_{\chi\chi}$ and $D_{rh}$
considered here as
\begin{equation}
R_{Dh}\simeq D_{rh}\left(\frac{\partial P_{\varphi}}{\partial\psi}\right)_{\mathcal{E},\chi}^{2}\left(\frac{\partial\psi}{\partial r}\right)^{2}/\left(\frac{\partial P_{\varphi}}{\partial\chi}\right)_{\psi,\mathcal{E}}^{2}\nu_{\chi\chi}=\frac{D_{rh}\left(\frac{e}{2\pi mc}\frac{\partial\psi}{\partial r}\right)^{2}}{\nu_{\chi\chi}\left(\frac{vRB_{\varphi}}{B}\right)^{2}}=\frac{D_{rh}}{\nu_{\chi\chi}\left(\frac{qR\rho_{h}}{r}\right)^{2}}.\label{eq:RDh}
\end{equation}
. Here the expression for $R_{Dh}$ is the same as in Ref.\citep{LangFu2011}
except that we do not rely on large poloidal mode number and define
the Larmor radius variable as $\rho_{h}=v/\omega_{c}$.

The first term on the RHS of Eqs.(\ref{eq:RBQDiffOp},\ref{eq:RBQDiff1D})
has AE driven diffusion coefficient which is coming primarily from
the window resonance function $\mathcal{R}_{\bm{l}}$ dependence on
the canonical momentum \citep{DuartePoP19}. The second term in Eq.(\ref{eq:RBQDiff1D})
is responsible for the pitch angle scattering which could be locally
dominated by either Coulomb collisions (when $R_{Dh}\ll1$) or microturbulence
(when $R_{Dh}\gg1$) induced radial diffusion since it has complicated
dependencies in the phase space, $D_{rh}=D_{rh}\left(P_{\varphi},\mathcal{E},\mu\right)$
\citep{ZhangLinChen2008PRL,Hauff2009PRL}. Eq.(\ref{eq:RBQDiff1D})
can be rewritten in formal action variables with the resonant frequency
$\Omega$ being a function of three constants of the unperturbed motion
\citep{BerkPPR97}: $P_{\varphi}$, magnetic moment, $\mu$, and energy,
$\mathcal{E}$, which extended the collisionless QL theory originally
developed by Kaufman \citep{KaufmanQLPoF1972}. We can then rewrite
it in the form 
\begin{equation}
\frac{\partial f}{\partial t}=\frac{\pi}{2}\sum_{k,p,m,m'}\frac{\partial}{\partial\Omega_{kp}}\left|\omega_{b}^{2}\right|^{2}\mathcal{R}\left(\Omega_{kp}\right)\frac{\partial}{\partial\Omega_{kp}}f+\nu_{eff}^{3}\frac{\partial^{2}(f-f_{0})}{\partial\Omega_{kp}^{2}},\label{eq:RBQGwGnsc}
\end{equation}
where 
\begin{equation}
\nu_{eff}^{3}\simeq\left[R_{Dh}+1\right]\left(\frac{\partial\Omega}{\partial\chi}\right)^{2}\nu_{\chi\chi},\label{eq:RDhPls1}
\end{equation}
and the window, or resonance, function $\mathcal{R}$ replaces the
resonance $\delta$-function and automatically satisfies $\int_{-\infty}^{\infty}\mathcal{R}\left(\Omega\right)d\Omega=1$
\citep{DuartePoP19}.

Note that in the RBQ quasilinear methodology used here the resonance
function is broadened over a characteristic width of $\Delta\Omega\simeq2.58\nu_{eff}$
\citep{DuartePoP19} whereas in BOT \citep{Lilley2010,LilleyBOT14}
it is computed using the kinetic equation. Even though the RBQ one-dimensional
results in comparison with the experimental data were favorable \citep{GorelenkovPoP19},
they lacked AE amplitude steady state saturation. The pitch-angle
scattering in those simulations was due to classical Coulomb collisions
and the used diffusion rate was taken at a time of maximum amplitudes.

\textbf{\textit{Comparison between RBQ and BOT simulations}}\textbf{.
}Here we compare RBQ simulations in tokamak geometry with fully nonlinear
BOT results obtained in a 1D model geometry where the kinetic equation
solution scheme resolves the structures near one resonance in the
Fourier space. A comparison between the QL methodology in a model
geometry \citep{GhantousPoP14} and BOT has been performed using a
heuristic broadening function has shown that although QL and BOT simulations
can exhibit fair qualitative agreement, quantitatively the agreement
only occurs in a limited parameter range. BOT and RBQ comparison is
done as close as possible by using the same input parameter of growth
and damping rates and effective pitch angle scattering.

\begin{figure}
\begin{centering}
\includegraphics[viewport=0mm 0bp 520bp 500bp,clip,width=8cm,height=5cm]{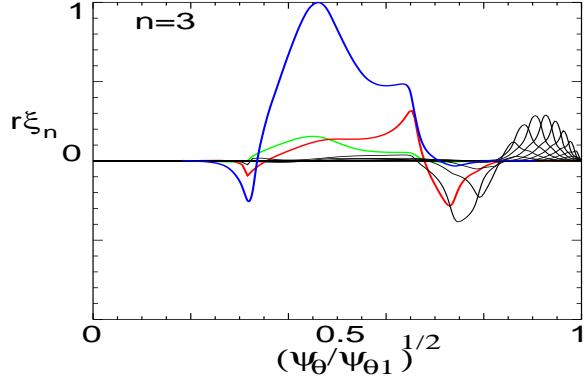}\emph{\small{}\vspace{-6mm}}{\small\par}
\par\end{centering}
\caption{\emph{\small{}Radial mode structures of n=3 RSAE poloidal harmonics
of radial plasma displacement vs. minor radius variable, $r/a=\sqrt{\bar{\psi}_{\theta}}$.
\vspace{-4mm} }\label{fig:AEn3}}
\end{figure}

In RBQ simulations, we consider one reversed-shear Alfvén eigenmode
(RSAE) with toroidal number $n=3$ corresponding to the observed unstable
mode at $t=805msec$ of DIII-D discharge $\#159243$ described in
Refs. \citep{CollinsPRL16}, which is one of the modes extensively
analyzed recently in several publications \citep{Taimourzadeh2019,GorelenkovPoP19}.
Its mode structure computed by the ideal MHD code NOVA \citep{ChengPF86}
is shown in Fig. \ref{fig:AEn3} as RSAE poloidal harmonics of $\boldsymbol{\xi}\cdot\nabla\psi_{\theta}$
radial dependence versus the minor radius flux variable. 

\begin{figure}
\begin{raggedright}
\hspace{30pt}\raisebox{110pt}{\normalsize (a)}{\small{}\hspace{-50pt}\includegraphics[viewport=0bp 5bp 660bp 280bp,clip,width=12cm,height=5cm]{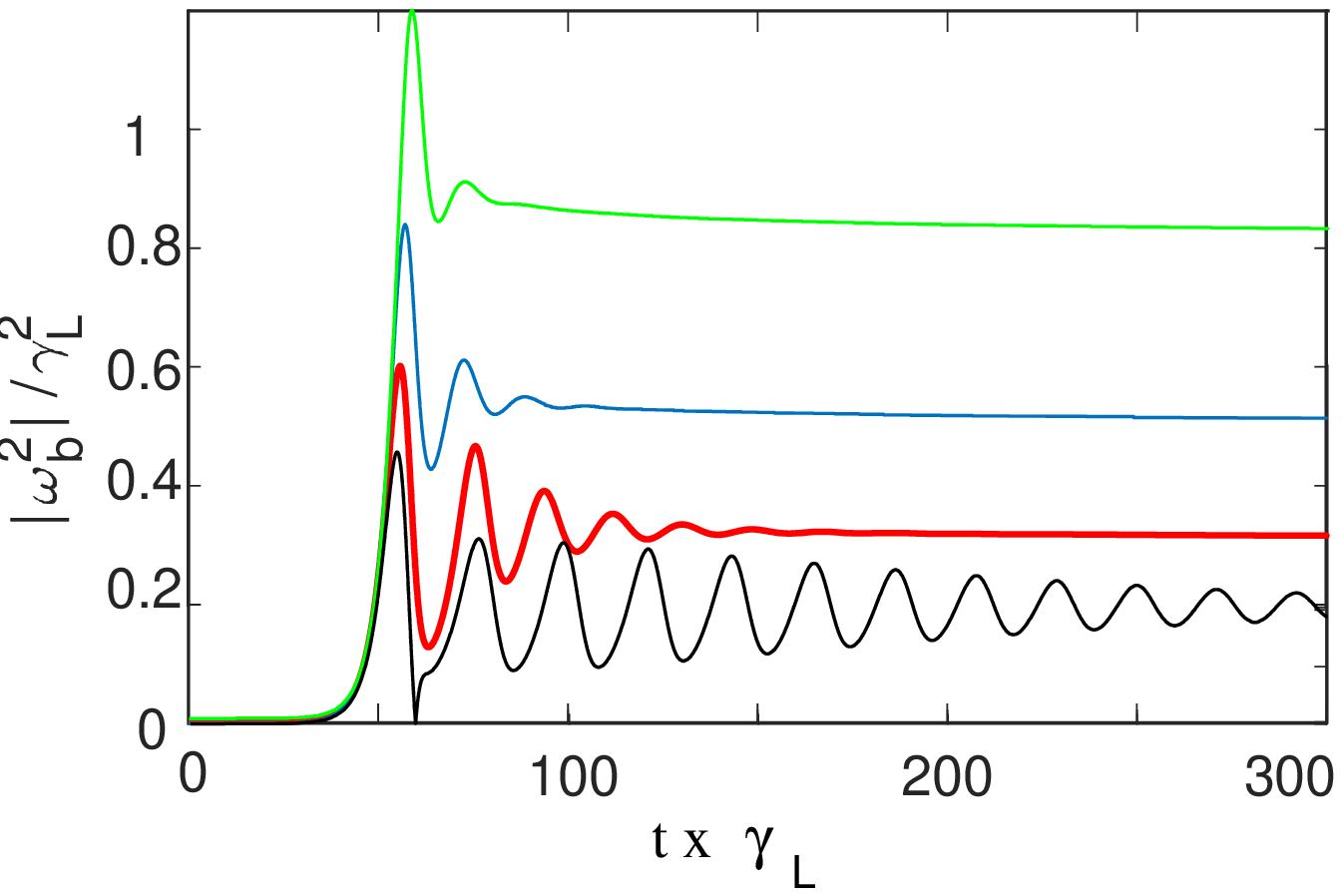}}\hspace{40pt}\raisebox{110pt}{\normalsize (b)}\hspace{-180pt}{\small{}\includegraphics[viewport=0bp -15bp 660bp 313bp,clip,width=12cm,height=5cm]{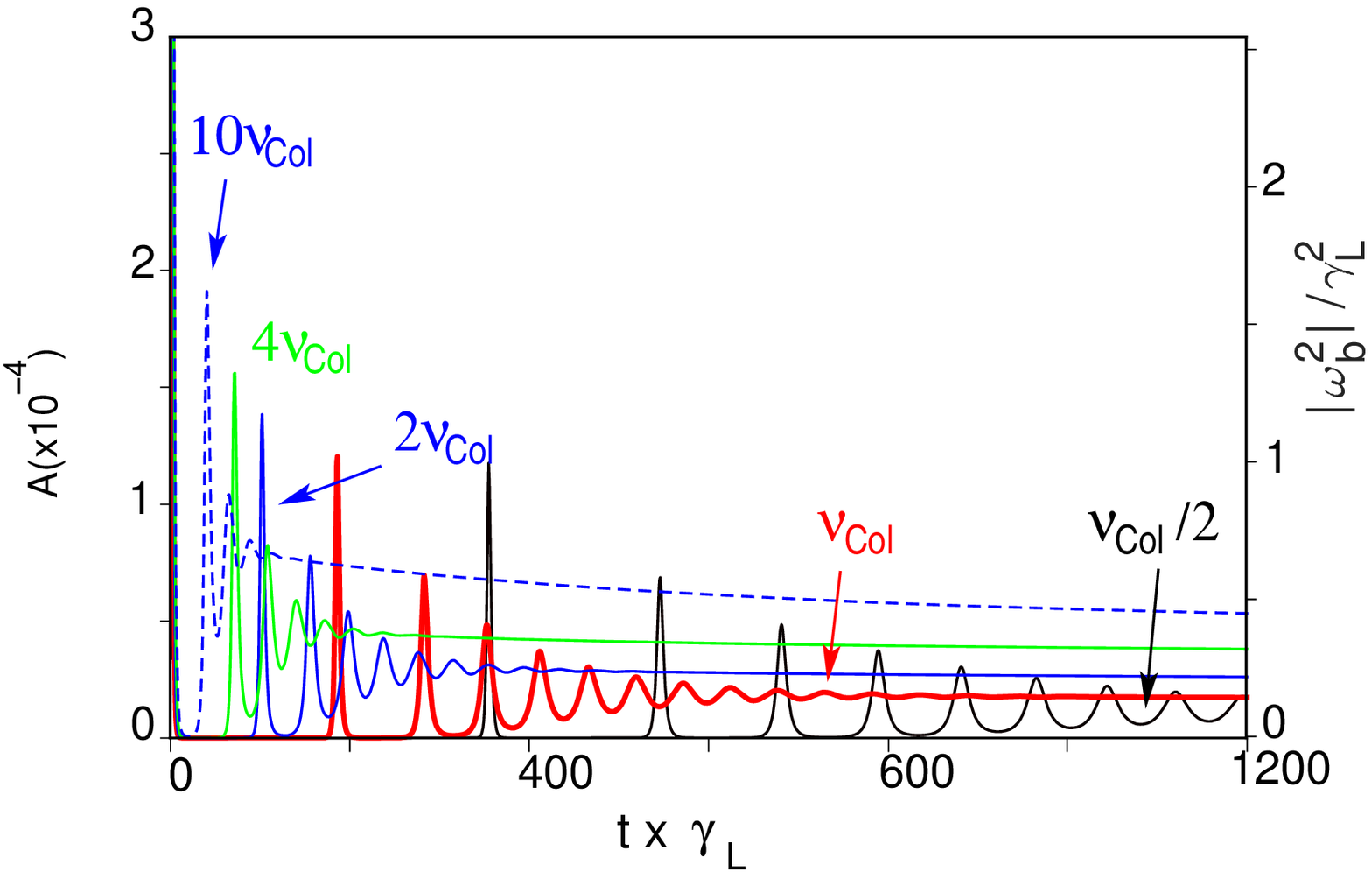}\hspace*{-100pt}}{\small\par}
\par\end{raggedright}
\caption{\emph{\small{}AE amplitude vs time from RBQ1D and BOT for different
degrees of collisionality. Left figure (BOT) has the effective frequency
rates (going from the bottom figure up) $0.49\gamma_{L},0.618\gamma_{L},0.778\gamma_{L}$
and $0.98\gamma_{L}$ ($\gamma_{L}$ is an input parameter of BOT).
They correspond to the RBQ scattering rates $\nu_{Col}/2,\nu_{Col},2\nu_{Col}$
and $4\nu_{Col}$ of the nominal scattering frequency $\nu_{Col}=8.9sec^{-1}$
computed by NOVA-K (right figure). Figures a and b have the same color
coding for the corresponding scattering frequencies, i.e. the red
curve is the nominal (collisional) scattering frequency. We also plot
a much larger value of the scattering frequency curve, $10\nu_{Col}$
for RBQ simulations as blue dashed like. \label{fig:Avst}}}
\end{figure}

We have found that if the $\chi$ scattering is given by Coulomb collisions,
$\nu_{\chi\chi}$, RBQ simulations (see Fig.\ref{fig:Avst} b) lead
to the overshoot (first maximum in time) point with quick amplitude
decaying with the damping rate, $\gamma_{d}<0$. The growing phase
time is determined by the $\gamma_{L}+\gamma_{d}>0$ rate. In the
collisional case the effective particle source due to the scattering
operator of Eq.(\ref{eq:RBQDiffOp}) is sufficiently weak to replenish
the resonant ions near the resonant region and the cycle does not
repeat at later times. However if the scattering is set up to a larger
value the fast ion population in the resonance region is replenished
and a new growing phase emerges. In a later case a classical predator-prey
interplay in the AE nonlinear dynamics outlined in Ref. \citep{BerkPRL92}
occurs. We should note that even though our model includes EP source
through the pitch angle scattering only, because of the time separation
energy slowing down contribution is much weaker (or slower) than the
scattering in pitch angles.

\begin{figure}
\includegraphics[width=8cm,height=6cm]{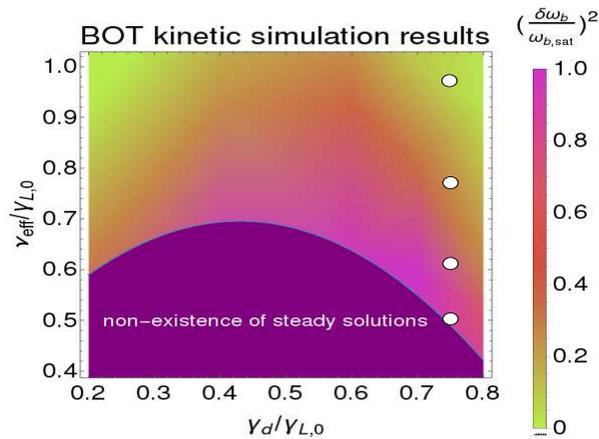}

\caption{\emph{\small{}Contour map of AE oscillations normalized by the saturation
amplitude computed by BOT simulations in units of the bounce frequency
at saturation $\omega_{b,sat}$. The white dots indicate the parameters
used in Fig. \ref{fig:Avst}. In the purple region, the solutions
have a pulsating amplitude pattern that prevents a steady state to
be achieved. \label{fig:ContMap}}}
\end{figure}

\label{radiusdependence}Within the QL methodology the interplay between
background damping and the scattering frequency controls the repetition
rate for AE peaks. These results, shown in figure \ref{fig:Avst}
b, are consistent with BOT shown in Fig. \ref{fig:Avst} a. Both figures
are plotted for the same scattering rate values indicated on the contour
map, Fig.\ref{fig:ContMap} as white circles. They correspond to the
nominal scattering frequency, $\nu_{eff}=8.017\,10^{3}sec^{-1}$ or
$\nu_{Col}=\left[R_{Dh}+1\right]\left(\frac{\partial\bar{P}_{\varphi}}{\partial\chi}\right)^{2}\nu_{\chi\chi}=8.9sec^{-1}$
computed by the NOVA-K code but with fixed value independent on the
minor radius. The sequence of used points (going up vertically) on
Fig.\ref{fig:ContMap}(a) are indicated. Used value in BOT, $\nu_{eff}=0.618\gamma_{L}$
(nominal, red curves), correspond to RBQ scattering $\nu_{Col}$ and
the same rate, $\gamma_{L}=1.3\,10^{4}sec^{-1}$. \label{Within-the-QL}

In comparison with previous model studies \citep{GhantousPoP14},
we show that the kinetic simulations of BOT agrees much better with
RBQ for the oscillatory behavior of the Alfvénic modes. This is illustrated
on Fig.\ref{fig:Avst} and is due to the fact that the resonance function
used in RBQ is derived self-consistently \citep{DuartePoP19}.

There most important difference between two simulations is the recovery
time between the peaks is about 30 to 50\% larger in RBQ for the same
scattering frequency than in BOT. This is because the coefficient
$\left(\partial P_{\varphi}/\partial\chi\right)^{2}$ in Eq.(\ref{eq:RBQDiff1D})
is proportional to $v^{2}$ and with the same $\nu_{\chi\chi}$ and
estimates for the resonance velocity in RBQ (as shown below), we find
larger time for the resonant particles and thus the recovery rate.
What comes as a surprise is that the experimental point lies near
the threshold of the existence and non-existence of steady state regimes
in both RBQ and BOT simulations whereas in DIII-D discharge of interest
the AE amplitudes are in a steady state regime \citep{CollinsPRL16}.
As discussed in Ref. \citep{GorelenkovPoP19} the boundary conditions
(BC) should not be physical in RBQ, which are reflective in the plasma
center, but they should be the fixed value BC to ensure the expected
analytic amplitude scaling $\delta B_{\theta}/B\sim\nu_{eff}^{2}$
\citep{BerkPRL92}.

A contour plot obtained with the BOT code is shown in Fig.\ref{fig:ContMap}.
One notable consistency between Figs. \ref{fig:Avst}, (a) and (b),
is the value of the normalized nonlinear bounce frequency, $\omega_{b}$,
coming out of RBQ and BOT simulations. Both models fairly agree with
each other when they are in near-threshold regimes.

\textbf{\emph{Microturbulence as an origin for anomalous beam ion
pitch angle scattering}}\emph{. }The pitch angle scattering considered
recently for the problem of AE frequency chirping \citep{DuarteNF17,DuartePoP17},
is proposed here as a mediator for the EP driven AE amplitude saturation.
The scattering can be expressed with the help of the canonical momentum
if the radial diffusion is known. However, the radial diffusion and
EP pitch angle scattering are difficult to evaluate without accurate
knowledge of the level of the microturbulence. Here we consider the
upper and lower bounds for the scattering frequency.

\label{Microturb}Let us first compare the Coulomb scattering frequency
with the scattering frequency resulting from the microturbulence acting
on fast ions. Two expressions for EP diffusion coefficient exist which
are projections from thermal ion heat conductivity up to the energetic
particle energies. The first one is by Angioni which is the diffusion
coefficient averaged over EP distribution \citep{AngioniNF09}, $D_{b}^{A}\simeq\frac{2}{3}\left(\chi_{i}+\chi_{e}\right)\left[0.02+4.5\frac{T_{e}}{E_{b0}}+8\frac{T_{e}^{2}}{E_{b0}^{2}}+350\frac{T_{e}^{3}}{E_{b0}^{3}}\right]$
where $E_{b0}$ is the injection energy. It includes the diffusion
transport produced by the electrostatic background plasma microturbulence.
It is computed using the quasilinear microturbulence models and fitted
with the help of several gyrokinetic codes. The second expression
results from the electrostatic GTC simulations \citep{ZhangLinChen2008PRL}
and, for passing particles, it is approximately $D_{r}=D_{r,i}\frac{5T_{i}}{E_{br}}=\frac{2}{3}\chi_{i}\frac{5T_{i}}{E_{br}},$
where $E_{br}$ is the fast ion energy at the resonance with the mode.
This expression was successfully validated within the chirping criterion
to DIII-D \citep{DuarteNF17,DuartePoP17}. Both expressions need to
be evaluated using realistic estimates for the mode frequency, which
is upshifted due to the finite plasma pressure \citep{VanzeelandNF16,GorelenkovPPCF06RSAE}:

\begin{equation}
\sqrt{\omega_{GAM}^{2}+\omega_{\nabla}^{2}+\omega_{AE}^{2}}\simeq\left|\left(k_{\|}\pm\frac{1}{qR}\right)\chi\right|v,\label{eq:resVparall}
\end{equation}
where $\omega_{GAM}$ is the geodesic acoustic modes (GAM) frequency,
$\omega_{\nabla}$ is the pressure gradient contribution to the frequency
shift, and $\omega_{AE}=k_{\|}v_{A}$ is the frequency of AE eigenmode
ignoring those effects. For classical TAEs $k_{\|}=1/2qR$, $\omega_{TAE}=v_{A}/2qR\simeq\sqrt{\omega_{GAM}^{2}+\omega_{\nabla}^{2}}$,
and one can get $\left|v_{\|}\right|=\left|v_{A},v_{A}/3\right|$
resonances from this equation if the GAM frequency is negligible.
In case of DIII-D, the RSAE mode upshift frequency can be small, near
the GAM value, $\omega_{RSAE}^{2}\ll\omega_{GAM}^{2}+\omega_{\nabla}^{2}$
and $\left|k_{\|}\right|\ll\frac{1}{qR}$. So that with good accuracy
we estimate for this case

\begin{equation}
\frac{v}{v_{A}}\simeq\frac{\sqrt{\omega_{GAM}^{2}+\omega_{\nabla}^{2}}}{2\omega_{TAE}\left|\chi\right|}\simeq\frac{1}{2\left|\chi\right|}.\label{eq:resVparallEstim}
\end{equation}
Alternatively, it can sweep up to the TAE frequency when the resonant
ion velocity goes down to $v/v_{A}\simeq\left(3\left|\chi\right|\right)^{-1}$.

\label{Rdhvalues}Extensive gyrokinetic simulations are required to
evaluate $R_{Dh}$ at each time of the discharge. Instead we infer
the electron and ion thermal conductivities from TRANSP simulations
for DIII-D shot $\#159243$ at $t=805msec$ to be $\chi_{e}=2.28m^{2}/sec$
and $\chi_{i}=1.27m^{2}/sec$. \label{AngEval}The above projections
of the thermal ion conductivity inferred from TRANSP modeling to EP
effective scattering rate provide the value of $R_{Dh}=0.15$ in case
of Ref.\citep{ZhangLinChen2008PRL} expression and $R_{Dh}=0.65$
in case of Ref.\citep{AngioniNF09} expression taken at $E_{br}=E_{0}$.
These estimates go in Eq.(\ref{eq:RDhPls1}) and are compared with
pure Coulomb collisional scattering $\nu_{Col}=8.9sec^{-1}$ when
$R_{Dh}=0$. More accurate gyrokinetic simulations, such as given
in Ref.\citep{DuarteNF18} are required.

We note thought that the above estimates imply that depending on the
injection angle, $E_{br}$ could be as low as $m_{h}\left(v_{A}^{2}/9\right)/2$
and as high as $E_{0}$.

There are other significant factors which need to be considered for
better estimates of $R_{Dh}$. The most important is the averaging
over the mode structure of thermal ion and electron conductivities.
TRANSP analysis shows that from $\left.q\right|_{\bar{\psi}_{q_{min}}^{1/2}=0.45}=q_{min}$
surface towards $\bar{\psi}^{1/2}=0.7$ where the RSAE structure is
bounded, electron and ion thermal conductivity approximately doubles.
Another factor is that both Angioni and Zhang's projections were done
in the electrostatic limit whereas the electromagnetic turbulence
\citep{Hauff2009PRL} was ignored. In our study we did not quantitatively
evaluate those effects. Our results indicate that the anomalous scattering
due to the microturbulence should be routinely included in experimental
interpretations and can be similar or stronger than the classical
Coulomb scattering.

\textbf{\emph{In conclusion}}\emph{,} we show that the classical Coulomb
collisions are too small to provide the pitch angle scattering and
to replenish the resonant ion population near the AE resonances. This
means that the unstable AEs can be in a saturated steady state regime
if additional scattering is involved. We show that to see such steady
state regimes one needs to include additional microturbulence induced
scattering which is expected to be 2-5 times stronger than the classical
Coulomb scattering. This conclusion provides an alternate route for
fast ion losses with regard to the arguments of Ref. \citep{Pace2013}.
We observe that microturbulence acts as a mediator of fast ion redistribution
by increasing the overall effective pitch angle scattering and thereby
increasing the level of saturation of AEs. This, in turn, leads to
enhanced Alfvénic transport that would not occur in the absence of
turbulence. Essential to our analysis is the conclusion of Ref. \citep{Duarte_NF2019}
that, sufficiently near marginal stability, the effect of scattering
collisions on a single resonance dynamics is to erase the system memory
so that quasilinear and nonlinear theories give the same governing
evolution equation for near-threshold instabilities. In comparison
with previous model studies \citep{GhantousPoP14}, RBQ (which uses
a resonance window function derived self-consistently from first principles
\citep{DuartePoP19}) has found much better agreement with the kinetic
simulations of BOT for the oscillatory behavior of the Alfvénic modes.
The route found to enhance EP redistribution is expected to significantly
enhance the Alfvénic mode driven effects on fusion alphas in ITER
plasmas \textcolor{black}{\citep{FitzgeraldNF16,SchnellerPPCF16},
which in-depth consideration is beyond the scope of this paper. However
in our earlier evaluations of effective pitch angle scattering \citep{DuarteNF18}
where Fig.9 illustrates approximately an order of magnitude stronger
scattering in the presence of micro-turbulence. We stress that the
intermittency of AE in the nonlinear regime as described by the QL
theory is justified by our comparison of RBQ and BOT simulations.}However
our analysis does not include the fast ion scattering and resonance
overlaps by other AEs, which also could have similar effect as the
microturbulence. Nonlinear wave-wave interaction also is not considered
in this work but could be important if the amplitudes becomes significant.
\begin{acknowledgments}
The authors appreciate suggestions made by H. L. Berk on the manuscript.
This work was supported by the US Department of Energy under contract
DE-AC02-09CH11466.
\end{acknowledgments}

\bibliographystyle{apsrev4-1}
\bibliography{fivep20prv1}

\end{document}